# External noise effects on the electron velocity fluctuations in semiconductors

D. PERSANO ADORNO[a], N. PIZZOLATO[ab] AND B. SPAGNOLO[ab]

[a]Dipartimento di Fisica e Tecnologie Relative and CNISM,

[b]Group of Interdisciplinary Physics

Viale delle Scienze, Ed. 18, Palermo, Italy

We investigate the modification of the intrinsic carrier noise spectral density induced in low-doped semiconductor materials by an external correlated noise source added to the driving high-frequency periodic electric field. A Monte Carlo approach is adopted to numerically solve the transport equation by considering all the possible scattering phenomena of the hot electrons in the medium. We show that the noise spectra are strongly affected by the intensity and the correlation time of the external random electric field. Moreover this random field can cause a suppression of the total noise power.

### 1. Introduction

The spectral density of the electron velocity fluctuations in a n-type GaAs bulk undergoes critical modifications under the action of two mixed high-frequency large-amplitude periodic electric fields [1-3]. In particular, recent studies based on Monte Carlo simulations have shown that the total power of the intrinsic noise is very sensitive to the amplitude and the frequency of the excitation signals and that wave-mixing conditions may give rise to a mere redistribution, an enhancement or a suppression of the noise level, depending on the constructive interplay between the two excitation fields. However, semiconductor based devices are always imbedded into a noisy environment that could strongly affect their performance. For this reason, to fully understand the complex scenario of the nonlinear phenomena involved in the devices response, we analyze the effects of the addition of an external random electric field in semiconductor electron dynamics.

An important study about the constructive aspects of noise and fluctuations in different non-linear systems has shown that the addition of external random perturbations to systems with intrinsic noise may affect the dynamics of the system in a counterintuitive way, resulting in a reduction of the total noise of the system [4]. The opportunity to reduce the diffusion noise in semiconductor bulk materials by adding a random fluctuating contribution to the driving static electric field has been tested in ref. [5]. In that paper the suppression of the total noise power in the presence of an external source of fluctuations, characterized by a Gaussian distribution and a characteristic correlation time, was investigated.

In the present work we analyze the modification of the intrinsic carrier noise spectral density induced by an external source of random perturbations in low-doped semiconductor materials driven by a high-frequency periodic electric field. The presence of the external noise modifies the electron average velocity and significantly affects the internal noise spectrum of the system. A Monte Carlo procedure is used to numerically solve the transport equation by keeping into account all the possible scattering phenomena of the hot electrons in the medium. We show that the addition of an external random source to the periodic driving electric field strongly affects the noise spectra and can cause a suppression of the total noise power.

## 2. Monte Carlo procedure and noise calculations

In this work, the motion of electrons in a GaAs semiconductor bulk in the alternating electric field is simulated by a Monte Carlo algorithm which follows the standard procedure described in Ref. [6]. The conduction bands of GaAs are represented by the $\Gamma$ valley, the four equivalent L-valleys and the three equivalent X-valleys. The parameters of the band structure and scattering mechanisms are also taken from Ref. [6]. Our computations include the effects of the intravalley and intervalley scattering of the electrons in multiple energy valleys and of the nonparabolicity of the band structure. Electron scatterings due to ionized impurities, acoustic and polar optical phonons in each valley as well as all intervalley transitions between the equivalent and non-equivalent valleys are accounted for. We assume field-independent scattering probabilities; accordingly, the influence of the external fields is only indirect through the field-modified electron velocities.

The semiconductor bulk is driven by an electric field with two components, a periodic and a random one:

$$E(t) = E_0 \cos(\omega t + \varphi) + \eta(t). \qquad (1)$$

In Eq.(1) the deterministic component brings the system in the resonant condition, by adopting an amplitude $E_0$=10 kV/cm and a frequency $\omega$=500 GHz, and the random electric field is modelled with an Ornstein Uhlenbeck (OU) stochastic process $\eta(t)$, which obeys the following stochastic differential equation:

$$\frac{\partial \eta(t)}{\partial t} = -\frac{\eta(t)}{\tau} + \sqrt{\frac{2D}{\tau}} \xi(t), \qquad (2)$$

where $\tau$ and $D$ are, respectively, the correlation time and the variance of the OU process, and $\xi(t)$ is the Gaussian white noise with the autocorrelation $<\xi(t)\xi(t')>=\delta(t-t')$. The OU correlation function is $<\eta(t)\eta(t')>=D\exp[-|t-t'|/\tau]$.

The changes of the intrinsic noise properties are investigated by the statistical analysis of the two-time symmetric autocorrelation function of the velocity fluctuations and of its mean spectral density, as described in Ref.s [7-9].

## 3. Results and conclusions

The starting point for our analysis is a direct measure of the changes in the spectral density of the electron velocity fluctuations caused by a randomly fluctuating periodic electric field. In Fig.1a we show how the spectral density is modified by the application of an electric field with a random fluctuating component of constant amplitude $D^{1/2}$=3 kV/cm, with respect to the case of a zero noise source ($\eta(t) = 0$), for two different values of the correlation time $\tau$. In Fig.1b the same plot is made by adopting a different amplitude $D^{1/2}$=5 kV/cm.

FIGURE 1 (a) and (b)

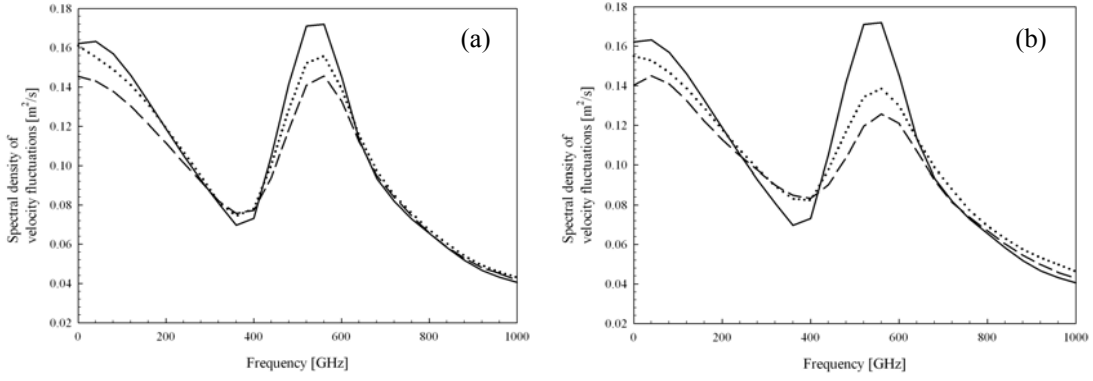

Figure 1: Spectral density of electron velocity fluctuations as a function of the frequency in the presence of an external source of noise with amplitude $D^{1/2}=3$ kV/cm (a) and $D^{1/2}=5$ kV/cm (b). Solid lines are used to describe the cases without any added noise; dotted and dashed lines show the spectral densities in the presence of noise with a correlation time $\tau_c=0.2$ ps (0.1 T) and $\tau_c=1$ ps (0.5 T), respectively.

In both cases (a and b), the spectral density of the electron velocity fluctuations appear to be significantly reduced by the introduction of the random term to the periodic electric field. How this effect of suppression is dependent on the amplitude and the correlation time of the OU noise is investigated by plotting the integrated spectral density (ISD) as a function of the correlation time (Fig. 2a) and the amplitude $D^{1/2}$ (Fig. 2b).

FIGURE 2 (a) and (b)

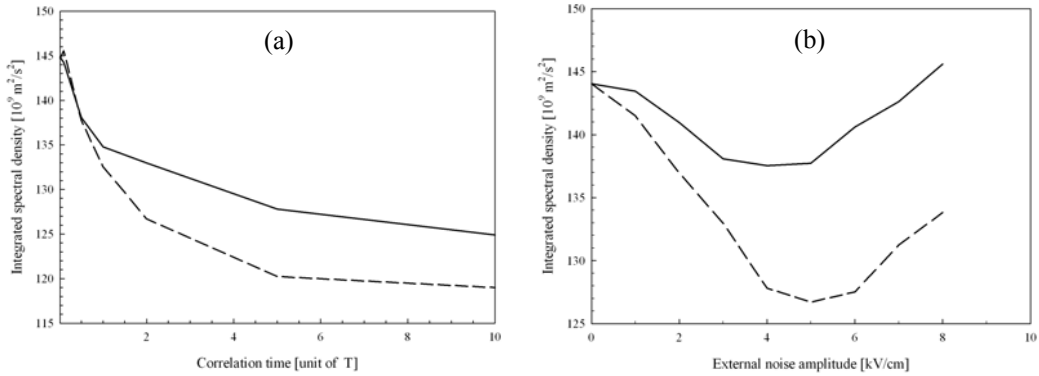

Figure 2: (a) Integrated Spectral Density (ISD) of electron velocity fluctuations as a function of the correlation time. Solid and dashed line is used for the noise amplitude $D^{1/2}=3$ kV/cm and $D^{1/2}=5$ kV/cm, respectively. (b) ISD of electron velocity fluctuations as a function of the external noise amplitude at fixed correlation time $\tau_c=0.5$ T (solid line) and $\tau_c=2$ T (dashed line).

The ISD, i. e. the total noise power, shows an exponential-like decreasing trend with the increasing of the correlation time for both values of the adopted noise amplitude (Fig. 2a). The suppression effect is more relevant in the case of $D^{1/2}=5$ kV/cm. In order to further clarify the role of the amplitude of the random fluctuations in the total noise power reduction, we have plotted the ISD as a function of the OU noise intensity in Fig. 2b, with two different correlation times, respectively 0.5 and 2 times the period $T=2\pi/\omega$ of the electric field. A very interesting nonlinear behaviour of the ISD with the noise intensity is found. In particular, we show the presence of a minimum with a position in the ISD vs. noise intensity diagram that critically depends on the correlation time.

The results reported in this work confirm that the total noise power in the presence of a high-frequency periodic and randomly-fluctuating electric field can be reduced. The transport dynamics of electrons in the semiconductor receives a benefit by the constructive interplay between the random fluctuating electric field and the intrinsic noise of the system, suggesting the existence of a stochastic resonance effect which could be responsible for the observed minimum of the ISD with the noise intensity.

Our future work will be oriented to the full comprehension of these phenomena and to the extension of this type of analysis to more complex structures.

## References

[1] M. C. Capizzo, D. Persano Adorno and M. Zarcone, *Phys. Stat. Sol. (c)* **3**, 2506 (2006).
[2] D. Persano Adorno, M.C. Capizzo and M. Zarcone, *J. Comput. Electron*, **5,** 475 (2006 ).
[3] D. Persano Adorno, M. C. Capizzo, and M. Zarcone, submitted to *Fluctuation and Noise Letters*.
[4] J. M. G. Vilar and J. M. Rubì, *Phys. Rev. Lett.*, **86**, 950 (2001)
[5] L. Varani, C. Palermo, C. De Vasconcelos, J. F. Millithaler, J. C. Vaissiere, J. P. Nouger, E. Starikov, P. Shiktorov and V. Gruzinskis, *Unsolved Problems of Noise and Fluctuations*, **474** (2005)
[6] D. Persano Adorno, M. Zarcone and G. Ferrante, *Laser Phys.* **10**, 310 (2000).
[7] P. Shiktorov, E. Starikov, V. Gružinskis, L. Reggiani, L. Varani and J. C. Vaissière, *Appl. Phys.Letters* **80**, 4759 (2002).
[8] P. Shiktorov, E. Starikov, V. Gruzinskis, S. Perez, T. Gonzalez, L. Reggiani, L. Varani and J. C. Vaissiere, *Phys. Rev. B* **67**, 165201 (2003).
[9] T. Gonzalez, S. Perez, E. Starikov , P. Shiktorov, V. Gružinskis, L. Reggiani, L. Varani and J. C. Vaissière in *Proceedings of SPIE* **5113**, 252 (2003).